\documentclass{article}

\usepackage{arxiv}

\usepackage[utf8]{inputenc} 
\usepackage[T1]{fontenc}    
\usepackage{hyperref}       
\usepackage{url}            
\usepackage{booktabs}       
\usepackage{amsfonts}       
\usepackage{nicefrac}       
\usepackage{biblatex}
\usepackage{doi}
\usepackage{orcidlink}
\usepackage{marvosym}
\usepackage{authblk}
\usepackage{graphicx}
\usepackage{bm}
\usepackage{amssymb}
\usepackage{gensymb}
\usepackage{tabularx}
\usepackage{caption}
\usepackage{subcaption}
\usepackage{algorithm}
\usepackage{algpseudocode}
\usepackage{amsmath}
\usepackage{multirow}
\usepackage{colortbl}

\begin{document}
\title{Perception and Control of Surfing in Virtual Reality using a 6-DoF Motion Platform}
\author[1]{Premankur Banerjee\orcidlink{0000-0002-0865-3634} \thanks{\href{mailto:premankur.banerjee@usc.edu}{premankur.banerjee@usc.edu}}}
\author[2]{Jason Cherin\orcidlink{0009-0001-0815-8756}}
\author[2]{Jayati Upadhyay\orcidlink{0009-0003-1207-5069}}
\author[2]{Jason Kutch\orcidlink{0000-0002-2417-4879}}
\author[1]{Heather Culbertson\orcidlink{0000-0002-9187-2706}}

\affil[1]{Thomas Lord Department of Computer Science, University of Southern California, Los Angeles, CA 90089, USA}
\affil[2]{Division of Biokinesiology and Physical Therapy, University of Southern California, Los Angeles, CA 90089, USA}

\date{}

\renewcommand{\headeright}{P. Banerjee et al.}
\renewcommand{\undertitle}{}
\renewcommand{\shorttitle}{Perception and Control of Surfing in VR}

\hypersetup{
pdftitle={Perception and Control of Surfing in Virtual Reality using a 6-DoF Motion Platform},
pdfauthor={Premankur Banerjee, Jason Cherin, Jayati Upadhyay, Jason Kutch, Heather Culbertson},
pdfkeywords={Virtual Reality, Motion Simulation, Acceleration Perception, Surfing},
}

\maketitle

\begin{abstract}
	The paper presents a system for simulating surfing in Virtual Reality (VR), emphasizing the recreation of aquatic motions and user-initiated propulsive forces using a 6-Degree of Freedom (DoF) motion platform. We present an algorithmic approach to accurately render surfboard kinematics and interactive paddling dynamics, validated through experimental evaluation with \(N=17\) participants. Results indicate that the system effectively reproduces various acceleration levels, the perception of which is independent of users' body posture. We additionally found that the presence of ocean ripples amplifies the perception of acceleration. This system aims to enhance the realism and interactivity of VR surfing, laying a foundation for future advancements in surf therapy and interactive aquatic VR experiences.
\end{abstract}

\keywords{Virtual Reality \and Motion Simulation \and Acceleration Perception \and Surfing}

\section{Introduction}
Virtual Reality (VR) simulations are becoming increasingly sophisticated with the addition of haptic feedback, striving to deliver experiences that closely mimic real-world activities. Aquatic simulations, particularly surfing, lag in their ability to create a high-fidelity and interactive simulation with appropriate kinesthetic feedback. 
Most simulators that integrate a 6-Degree of Freedom (DoF) motion platform have traditionally focused on air or ground vehicles~\cite{labedan2021virtual,colombet2009motion}. It is challenging to directly transfer these methods to aquatic simulations, especially for activities like surfing, because of the unique dynamics involved in water-based activities. In aquatic environments, the interaction with constantly changing water surfaces, wave dynamics, and the intricate maneuvering skills required are markedly different from the relatively stable and predictable environments of the air or ground. These complex motion dynamics make it difficult to accurately replicate the experience of surfing using simulation models initially designed for air or ground vehicles.
Additionally, case studies~\cite{riera2022case} and discussions on multimodal setups for non-aerial simulations~\cite{casas2017four} have highlighted the limitations of motion platforms in replicating the large-scale translational displacements expected in aquatic simulations. They also discuss the challenges faced by conventional Motion Cueing Algorithms (MCAs) and washout filters~\cite{liao2004novel} in accurately reproducing low level accelerations. MCAs are used to translate vehicle dynamics into perceptible physical motions on the motion platform, whereas washout filters, essential to MCAs, are designed to reset the simulator's position over time, preventing the motion platform from reaching its physical limits, while maintaining a realistic perception of motion. Since conventional MCAs are insufficient at replicating the large-scale motions in aquatic interactions, we must create a novel algorithm.
This paper introduces an approach of generating and mapping aquatic motions for surfing in VR, by integrating a 6-DoF motion platform to emulate surfboard dynamics and provide motion feedback.
\begin{figure}[t!]
    \centering
    \includegraphics[width=0.9\textwidth]{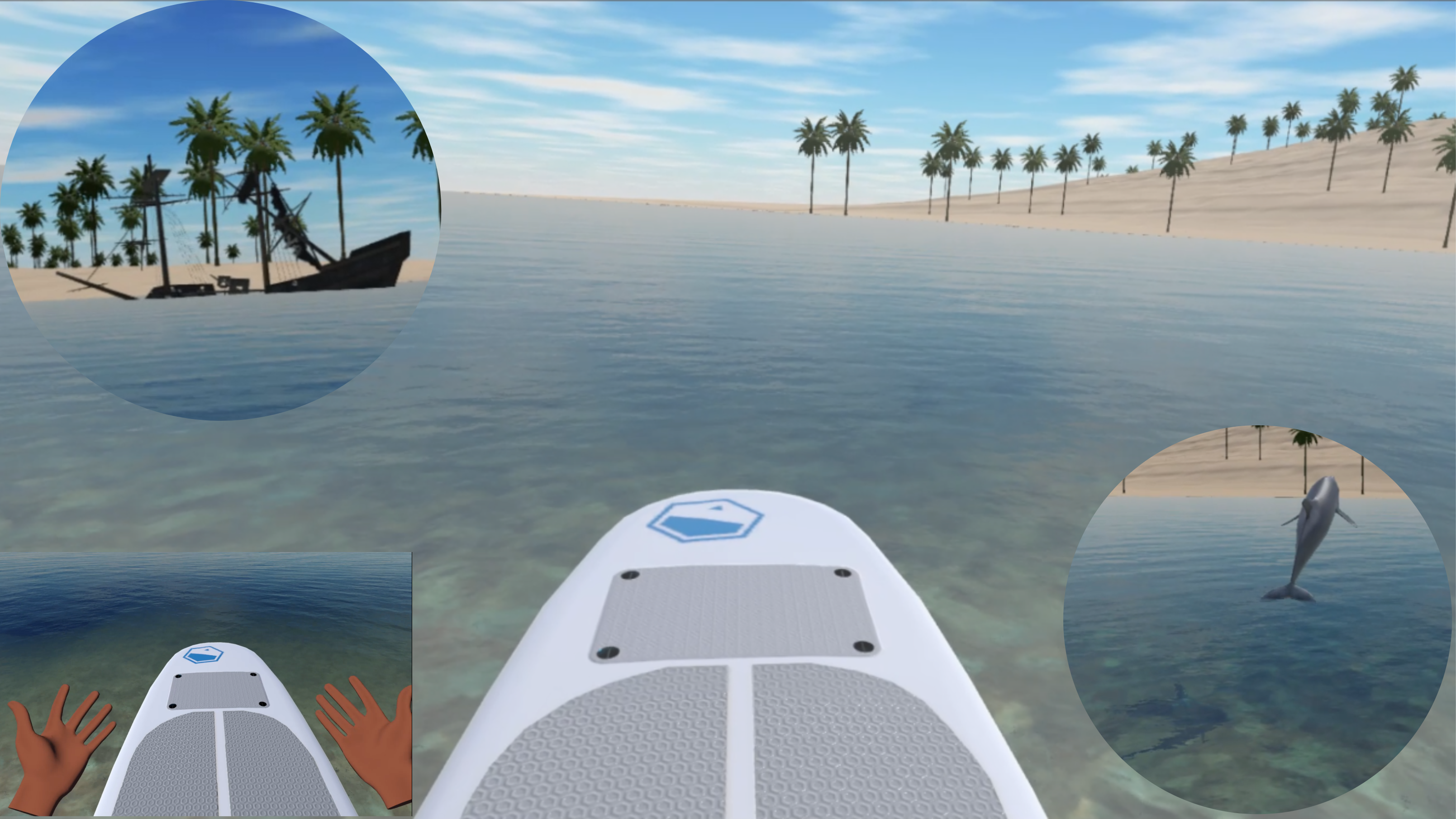}
    \caption{The VR Scene: PoV of user on surfboard with (inset) half-submerged pirate ship, pod of dolphins, and hands tracked and displayed by the headset}
    \label{VR_scene}
\end{figure}

Prior aquatic simulation research with motion platforms has predominantly focused on replicating the movements of boats or ships driven by external forces such as waves, wind, and currents. Ueng et al. noted that a ship's heave, pitch, and roll are primarily influenced by waves, whereas its surge, sway, and yaw are induced by external forces from propellers, rudders, and currents~\cite{ueng2008ship}. Subsequent studies have explored the effects of roll, pitch, and yaw on virtual sailing experiences~\cite{mulder2012effect}, leading to motion systems that enhance presence~\cite{mulder2013development}. Casas et al.\ introduced a simplified physics model to facilitate real-time speed-boat simulations on a 6-DoF motion platform~\cite{casas2012real}. 
While these simulators replicate externally driven motions, they do not account for user-generated forces (e.g., paddling or rowing), a critical aspect of activities like surfing~\cite{nessler2019electromyographic}. Creating a simulator that responds to user-generated motion presents unique challenges in accurately rendering low-magnitude accelerating forces while ensuring synchronicity with the user's motion. Researchers have proposed a rowing simulator with multimodal (visual, audio, and haptic) feedback that responds to user-generated input from oars~\cite{von2008real}. The above studies use either a screen or CAVE setup to display the virtual environment. Modern VR headsets offer a more versatile, cost-effective, and personalizable immersive experience compared to CAVE setups. To the best of our knowledge, there has not been any study that discusses motions on water due to user-generated propulsive forces by active paddling in VR.

In a recent study, a passive VR interaction was created with pre-recorded surfing videos to investigate VR's potential in enhancing the accessibility and appeal of surfing~\cite{huang2023surfing}. The study indicated that VR surfing appeals to a wide spectrum of users by providing a safe, controlled, and enjoyable environment, irrespective of users' geographic location. Surfing also holds promise as a therapeutic intervention by combining physical activity with a natural aquatic environment, offering benefits that extend beyond traditional therapies. Surf therapy has been shown to improve physical fitness, psychological well-being, and social interactions, particularly for populations like youth in need of social support, individuals with lower chronic back pain and other disabilities, and veterans~\cite{benninger2020surf}. 
Although positive results have been found for surf therapy in the ocean, it remains unstudied whether a VR surfing simulator would have similar therapeutic effects. However, a major benefit of VR is that it is more accessible than traditional surfing and would open up surf therapy to a larger group of people. This paper aims to explore the recreation of surfing motions and experiences in VR, positing that such an advancement could extend its applications to therapy.
\setcounter{footnote}{0}

Our paper aims to provide insights into the design considerations, algorithmic approach, and empirical validation of such a simulation system (see Fig. \ref{VR_scene}). Section 2 presents a mapping algorithm that estimates surfboard kinematics in VR and translates these into motion platform outputs. Section 3 details our experimental evaluation, which validates the system's effectiveness at creating a convincing and immersive surfing experience. We assess users' ability to discern different levels of aquatic accelerations and the impact of various body postures and presence of ocean ripples 
(small, rhythmic, wind-induced surface waves on the ocean, that induce a gentle, perceptible rocking motion on boats)
on the perception of acceleration.

\section{System Design and Control}
The following sections outline the techniques for mapping surfboard movements from VR to motion platform output, and the method for moving around within the simulation.

\subsection{Hardware and Software} Our setup consists of a 6-DoF motion platform (PS-6TM-150, \href{https://motionsystems.eu/}{MotionSystems}) to provide real-time haptic feedback of events occurring in the simulation. 
We securely mounted a surfboard to the motion platform, enabling users sit, kneel, or stand on the board. The system integrates a fan, shown in Fig.~\ref{fig:sub1}, that simulates wind by generating airflow at a continuous speed. The VR environment is rendered using a Meta Quest 2 headset. The simulation is designed in Unity 3D, using the \href{https://crest.readthedocs.io/en/stable/}{Crest} package (an open-source ocean renderer) to simulate waves. Within Crest, the waves' visual characteristics and dynamics are governed by a Fast Fourier Transform (FFT) spectrum, which can be manually manipulated or automatically generated. This spectrum offers adjustable parameters for different wave scales, alignment with wind direction, and horizontal displacement (chop), allowing for fine-tuned control. The virtual surfboard is designed to have both \texttt{Rigidbody} (physics-based ways to control movements via simulated forces and torque) and \texttt{WaterObject} (buoyancy and hydrodynamics calculations) components. The Crest package and the physics solver account for the motions induced by user-generated forces, as well as those generated by current, waves, and wind.

\subsection{Motion Mapping Algorithm} Our algorithmic approach provides a real-time estimation of the the surfboard kinematics within a Unity simulation and its corresponding mapping to our motion platform. Contrary to the operational dynamics of vehicle or flight simulators where external forces govern motion, a surfboard is driven by user-generated propulsive forces. Active paddling produces reactive forces that result in forward acceleration of the surfboard. Our algorithm seeks to render these low-magnitude accelerating forces with high fidelity, ensuring the resultant motion platform output is both realistic and synchronous with the user's motion inputs. 
\begin{figure}[tb]
    \centering
    \begin{subfigure}[b]{\textwidth}
        \includegraphics[width=\linewidth]{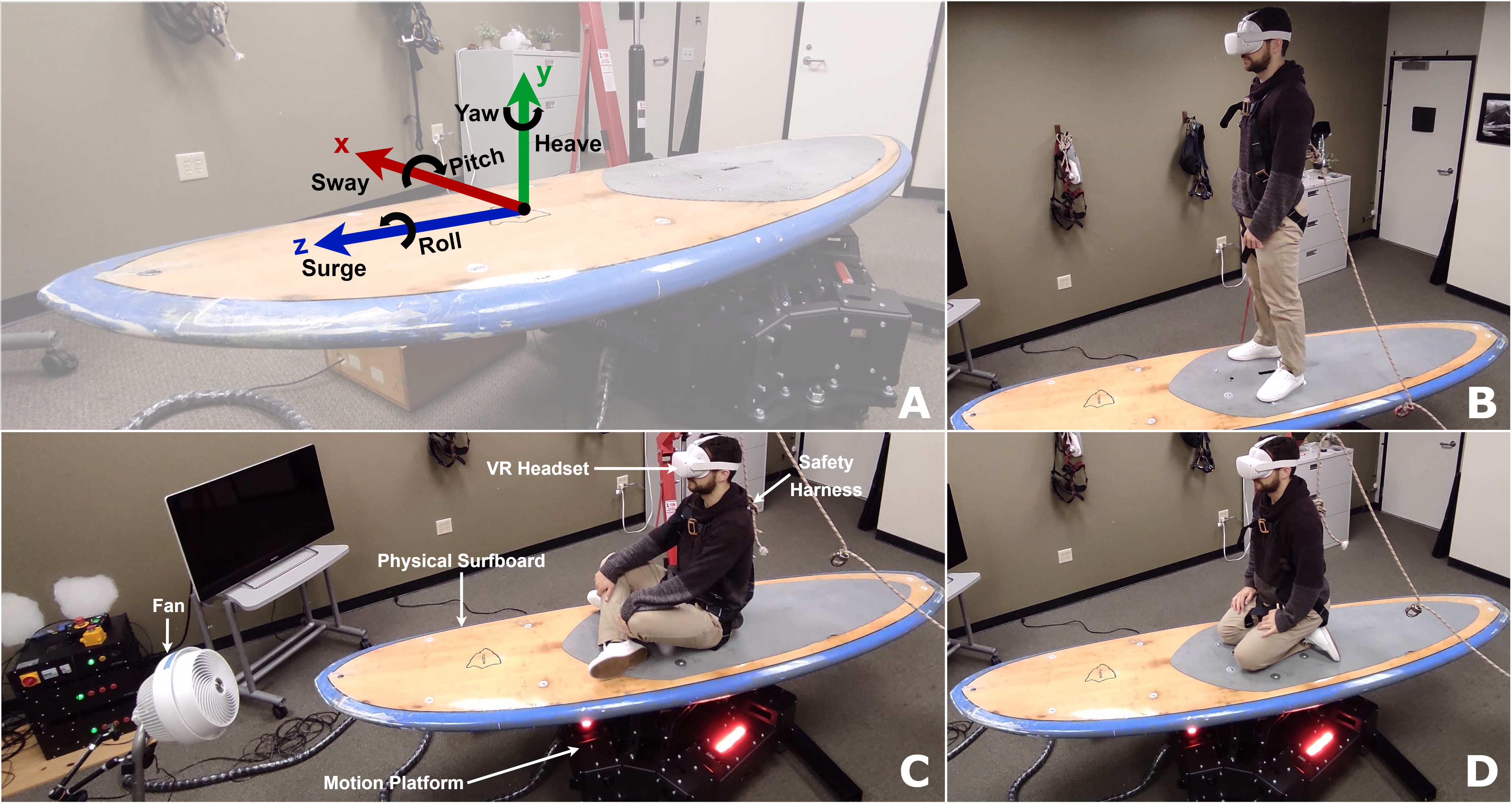}
        \caption{A) Coordinate Frame, B) Standing, C) Sitting; and Experimental Setup, D) Kneeling}
        \label{fig:sub1}
    \end{subfigure}
    \vspace{1mm} 
    \begin{subfigure}[b]{\textwidth}
        \includegraphics[width=\linewidth]{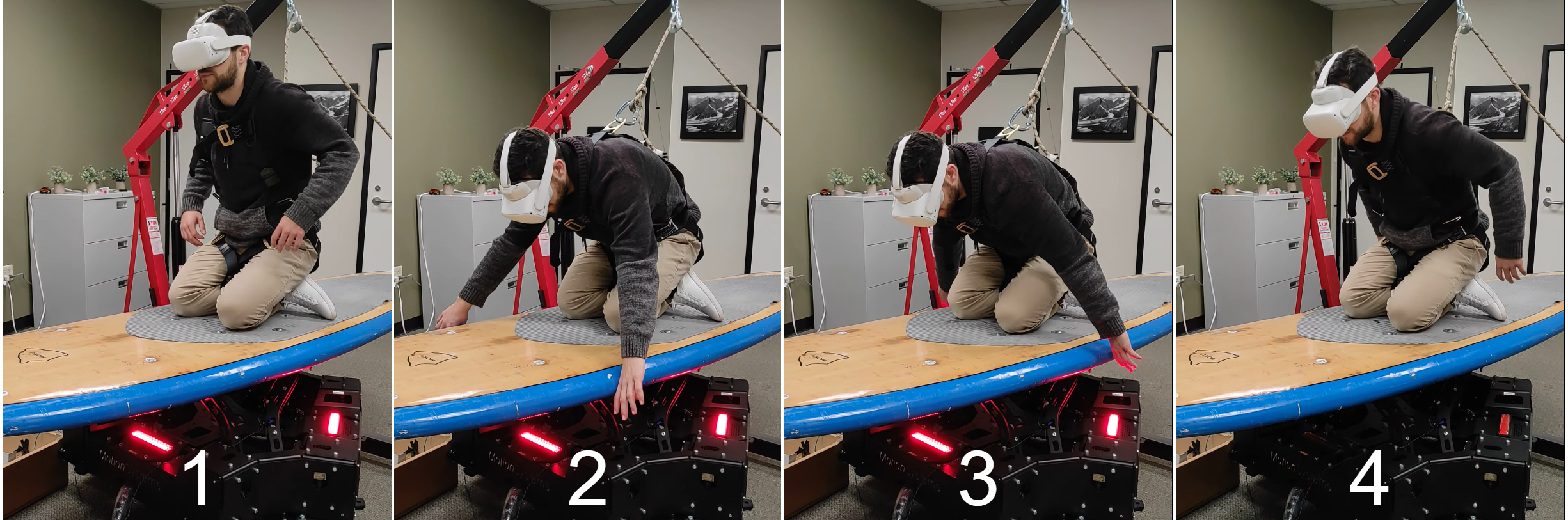}
        \caption{1) Preparing to paddle, 2) and 3) Immersing hands in virtual water, and executing a stroke (Propulsion) 4) Lifting hands out for the next stroke (Recovery)}
        \label{fig:sub2}
    \end{subfigure}
    \caption{Setup with three body postures (above) and Paddling in VR (below)}
    \label{fig:test}
\end{figure}

\begin{figure}[t!]
    \centering
    \includegraphics[width=\textwidth]{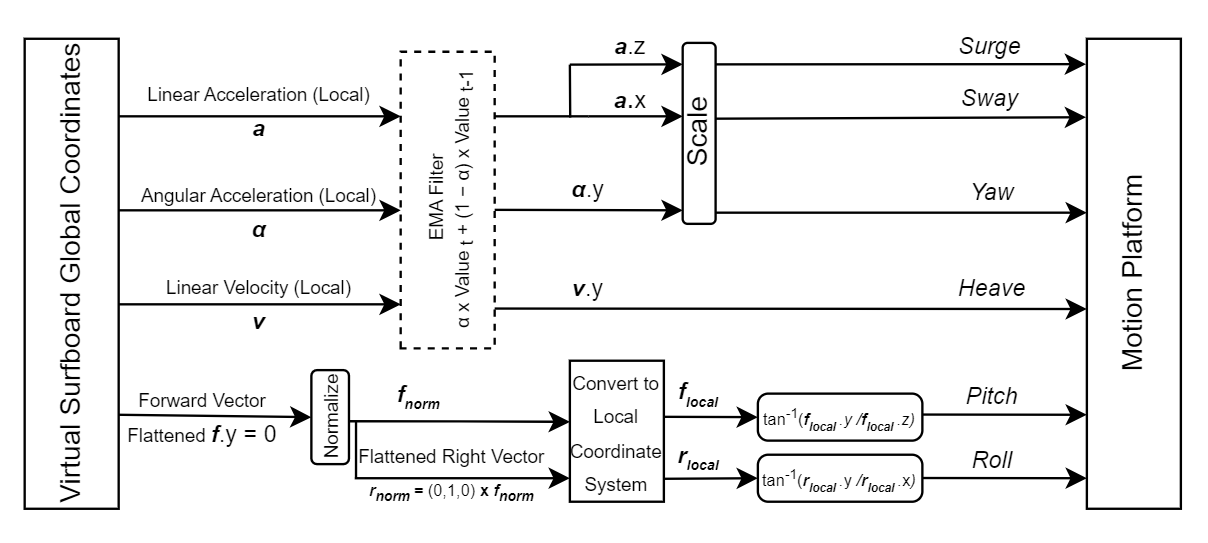}
    \caption{Flow Diagram of the system's Mapping Algorithm}
    \label{flowdiagram}
\end{figure}

The surfboard's acceleration (\(\Vec{a}\)) 
is computed by temporal differentiation of its velocity vector (\(\Vec{v}\)).
This raw acceleration vector (\( \Vec{a} = \Delta \Vec{v} / \Delta t\)) is then transformed into the surfboard's local reference frame, aligning it with its intrinsic axes. An Exponential Moving Average (EMA) filter is applied to the acceleration values (\(\Vec{a}_{filtered} = \lambda \cdot \Vec{a}_{t} + (1-\lambda) \cdot \Vec{a}_{t-1}\), where $\lambda$ is the filter coefficient) to produce a smoothed representation of the surfboard's motion~\cite{SalterAutomatedVehicles}.
We chose an EMA filter due to its rapid response to high-frequency, transient kinematic inputs, which facilitates precise rendering of real-time feedback necessary for the simulation of wave-induced motions and paddling in water.
For low acceleration values, the EMA ensures that the motion platform's output closely mirrors the surfboard's instantaneous kinematic state, providing a realistic experience. This is in contrast to the washout filter, a popular choice for simulations using motion platforms~\cite{liao2004novel}, which tends to dampen and delay signals and is sub-optimal in the context of surfing simulation, where high-fidelity replication of wave dynamics and board response is essential~\cite{casas2017four}. The washout filter could impair the immersiveness of the simulation, particularly in scenarios that require active paddling or navigating through turbulent waves. The filter's design to counteract actuator drift and maintain a neutral platform state does not align with the intrinsically dynamic nature of surfing, where sustained g-forces are not a primary concern.
\paragraph{\textbf{Acceleration-based Mapping}:}The motion platform's 6 DoF are constrained within predefined operational bounds. The kinematic parameters surge, sway, and yaw -- corresponding to the surfboard's longitudinal (z-axis), lateral (x-axis), and vertical axis rotations ($\phi$), respectively (see Fig.~\ref{fig:sub1}) -- are extrapolated from the virtual surfboard's linear and angular accelerations. Since these parameters are inherently unbounded in VR we cannot directly map them to the motion platform due to its operational bounds in these directions. Instead, we map filtered and scaled versions of the accelerations along these vectors, which exceed human perceptual thresholds~\cite{LeeVehicleEnvironments}, to the motion platform's actuators (Fig. \ref{flowdiagram}):
\begin{equation}
    \textrm{Surge}=SF_1\cdot \Vec{a}_f.z
\end{equation}
\begin{equation}
    \textrm{Sway}=SF_2\cdot \Vec{a}_f.x
\end{equation}
\begin{equation}
    \phi = SF_3\cdot \Vec{\alpha}_f.y
\end{equation}
where $\Vec{a}_f$ and $\Vec{\alpha}_f$ are the filtered linear and angular acceleration vectors, and $\phi$ is the yaw  of the virtual surfboard. The scaling factors $SF_1$, $SF_2$, and $SF_3$ were empirically optimized by an initial pilot study involving 6 participants to emulate realistic surfboard dynamics.
\paragraph{\textbf{Velocity and Position-based Mapping}:}Conversely, heave (y-axis), pitch ($\theta$), and roll ($\psi$) are spatially limited within VR due to the surfboard's interaction with the water surface. Heave is restricted since the surfboard can only rise or fall within the limits set by the water level displacement and board buoyancy. Similarly, the angles at which it can tilt forward or backward (pitch) and side to side (roll) without losing stability or flipping over are limited.

Heave is regulated through a filtered velocity-based mapping protocol:
\begin{equation}
    \textrm{Heave}=\Vec{v}_f.y
\end{equation}
which is chosen because the vertical velocities encountered in the surfboard simulation largely do not exceed the motion platform's operational thresholds, removing the need for acceleration-based scaling. We avoid a position-based mapping since the water level height might change due to the onset of waves, thereby pushing beyond the operational thresholds of the motion platform.

Pitch and roll are mapped directly to their corresponding angular displacements in the virtual surfboard's local coordinate frame. The algorithm projects the surfboard's orientation vector onto the transverse plane, discarding the vertical component to isolate the horizontal forward vector, which is then normalized. This approach ensures rotational invariance. By adopting a body-centric frame, global coordinates are transformed into a local reference frame, allowing for the calculation of intrinsic angular displacements. 
To allow the motion platform to accurately mirror the surfboard's tilt, pitch and roll are calculated as:
\begin{equation}
    \theta = tan^{-1}\left ( \frac{\Vec{f}_{long}.y}{\Vec{f}_{long}.z} \right )
\end{equation}
\begin{equation}
    \psi = tan^{-1}\left (\frac{\Vec{r}_{lat}.y}{\Vec{r}_{lat}.x} \right)
\end{equation}
where $\theta$ is pitch, $\psi$ is roll, $\Vec{f}_{long}$ is the projection of the forward vector on the longitudinal axis, and $\Vec{r}_{lat}$ is the projection of the right vector on the lateral axis.

\subsection{\textbf{Navigating in VR}} The simulation replicates real-world hand paddling dynamics~\cite{nessler2019electromyographic}. Users first submerge their hands in the virtual water and perform a stroke motion during the propulsive phase. They then lift their hands out of the virtual water, marking the beginning of the recovery phase, which lasts until they submerge their hands again for the next stroke (see Fig.~\ref{fig:sub2}). To recreate this in VR, we use the in-built hand tracking feature of the Meta Quest 2 VR headset. This paddling technique allows the user to freely move in all directions in the virtual environment. 

Our algorithm detects when hands are below the virtual water surface and applies appropriate reaction force and torque to the surfboard based on hand velocity in the longitudinal (surge) and lateral (sway) directions. A linear force, propelling the surfboard forward, is generated by longitudinal hand movements, while a combination of longitudinal and lateral hand movements induce a torque around the vertical (heave) axis, enabling steering.
Forward motion is achieved through a front-to-back stroke motion executed by both hands, while backward motion results from the reverse hand movement. Steering is directional -- paddling with the right hand turns the surfboard left, and paddling with the left hand turns the surfboard right. The degree of steering and propulsion correlates with the hand's movement along the lateral and longitudinal axis, respectively. Magnitude of these forces and torques is proportional to the hands' velocity:
\begin{equation}
   \Vec{F} \propto -S\cdot \Vec{v}
\end{equation}
where $\Vec{v}$ is the combined directional velocity of left and right hands, and \(S\) is the scaling factor (determined empirically by a pilot study with 6 participants).
The drag force, which resists the motion of the surfboard, is accounted for by the Crest physics, \texttt{Rigidbody}, and \texttt{WaterObject} components in Unity, and does not need to be explicitly considered in the paddling equations. This interaction model ensures that a combination of hand movements in both longitudinal and lateral axes creates realistic surfboard acceleration and directional control. In the simulation, we standardize the combined weight of the user and surfboard to be 150 kg, disregarding individual weight variations, to maintain consistency as they navigate through the environment. We also do not take into account the various hand sizes (surface areas) or depth of hand immersion in water (along vertical axis) which affect the propulsive or steering forces in real life. The forces in our implementation are solely dependent on the velocity of the hands.

\section{Experimental Evaluation}

We evaluate our proposed mapping algorithm and VR navigation via two experiments. Specifically, we evaluate the mapping algorithm on the basis of realism of interaction and users' perception of acceleration. During the experiment, we vary the level of acceleration, the user's body posture, and the presence of ocean ripples. Our hypotheses were: \textbf{H1.} Users can distinguish between different levels of accelerations; \textbf{H2.} Body postures (sitting, kneeling or standing) will affect acceleration perception; \textbf{H3.} Ocean ripples will affect acceleration perception; \textbf{H4.} The movements generated by the motion platform will feel realistic and congruent to those of the virtual surfboard.
The three accelerations provided to the virtual surfboard for the experiment testing \textbf{H1} to \textbf{H3} were 0.5, 1.5, and 3 $m/s^{2}$, which were classified as Low (LA), Medium (MA), and High (HA), respectively. LA was selected to align with the linear acceleration typically experienced during hand paddling in water~\cite{nessler2019electromyographic}. The magnitude of HA ($\approx$ 6$\times$LA) was representative of the linear acceleration encountered when being propelled by a wave or when being towed by a jet-ski~\cite{casas2013characterization,good2005stopping}. MA was assigned a value in between the two ($\approx$ 3$\times$LA). Ocean ripples were simulated by tuning the FFT spectrum used to simulate oceans from the Unity Crest package. These ripples are small, rhythmic, wind-induced surface waves on the ocean, that induce a gentle, perceptible rocking motion on boats.
\begin{figure}[t!]
    \centering
    \includegraphics[width=\textwidth]{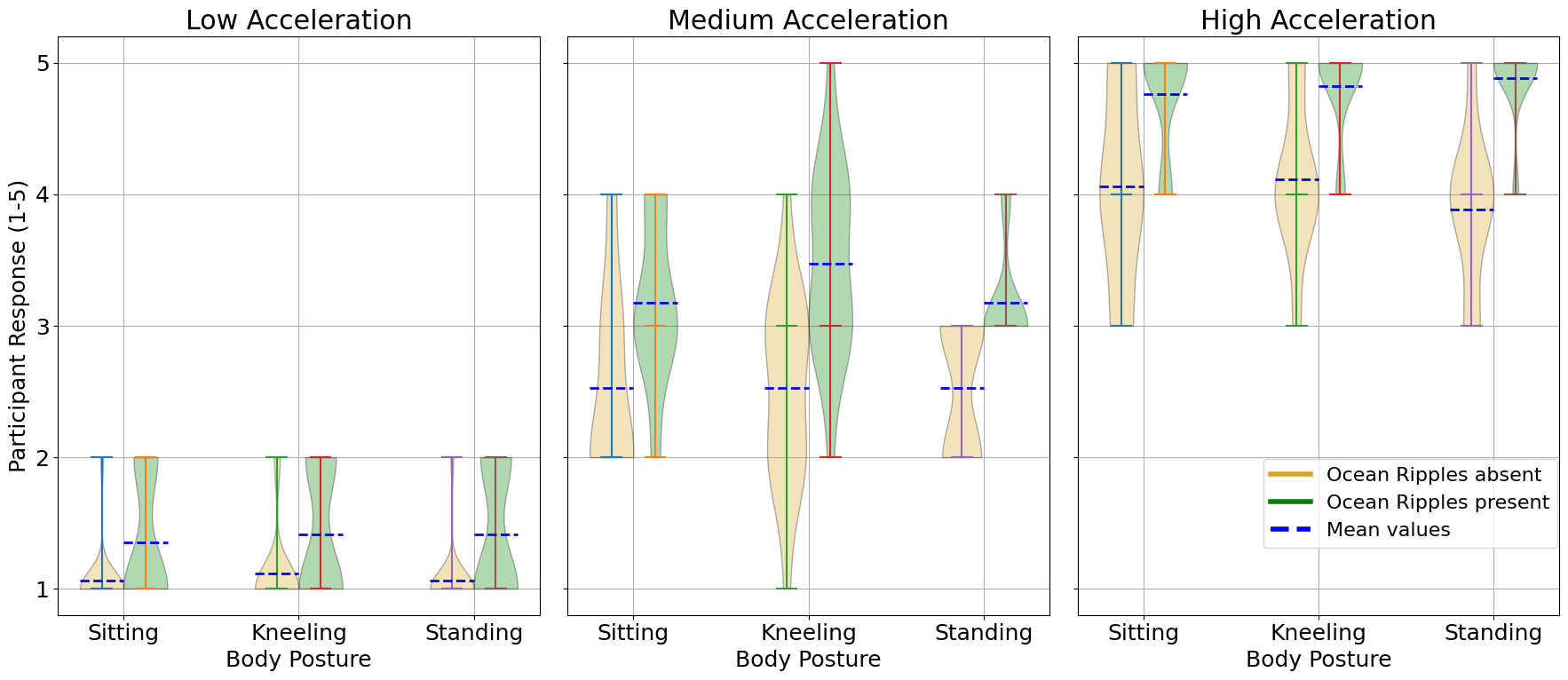}
    \caption{Violinplot of N=17 participant responses on a 5-point Likert scale to Acceleration Levels with and without Ocean Ripples across various Body Postures}
    \label{responseFig}
\end{figure}
\paragraph{\textbf{Participants}:} Seventeen participants (9 females, 8 males, $M_{age} = 27.59 \pm 3.28$) were recruited for the study. Ten had prior surfing experience and fourteen had prior VR experience. None had any sensory or motor impairments. The study was approved under USC IRB Protocol HS-23-00077, and all participants gave informed consent.
Before starting, participants were trained on how to move the surfboard around in VR by paddling. Hardware constraints of the motion platform plus a harness ensured participant safety. Participants wore noise-cancelling earphones connected to the VR headset which, in addition to cancelling noise from the motion platform, also played audio from the VR simulation. 

\subsection{Experiment 1: Perception of Aquatic Acceleration} \label{Exp1}
Experiment 1 examined participants' ability to distinguish three distinct levels of forward acceleration on a simulated ocean surface. The study further investigated the impact of body posture and ocean ripples on acceleration perception.

\paragraph{Procedure:} 
Participants were positioned on a surfboard in three distinct body postures: sitting, kneeling, and standing. For each posture, they were received three acceleration levels, once in the presence of ocean ripples and once without, for a total of 18 trials per participant. In each trial, participants experienced a forward linear acceleration for a duration of 5 seconds. 
Starting from a stationary position, the surfboard was accelerated for 5 seconds, then decelerated back to rest.
Participants received visual cues through the headset and kinesthetic cues from the motion platform. Post-trial, they quantified their \textit{perceived acceleration intensity} on a Likert scale ranging from 1 (very low) to 5 (very high).
Each participant was instructed to consistently maintain each of the three distinct body postures during their respective trials. For each posture, the presence of ocean ripples and acceleration levels were pseudo-randomized using a Graeco-Latin Square design. Additionally, the sequence in which the body postures were presented to the participants was randomized to preclude any potential order effects. The experiment lasted approximately 25 minutes per participant.

\paragraph{Results:} 

A factorial ANOVA assessed the effects of body posture, presence of ocean ripples, and acceleration levels on the perceived acceleration. Significant effects were found for acceleration levels \(F(2, 288) = 916.5, p < 0.001, \eta^2 = 0.817\) and ocean ripples \(F(1, 288) = 104.3, p < 0.001, \eta^2 = 0.046\), but not for body posture \(F(2, 288) = 0.936, p = 0.393, \eta^2 = 0.00084\). 
The interaction between ocean ripples and acceleration levels was significant \(F(2, 288) = 6.46, p = 0.0018, \eta^2 = 0.0058\). Other interaction effects were non-significant (\(p>0.05\)).
Post-hoc analyses using Tukey's HSD test revealed significant differences between acceleration levels: HA vs. LA (\(\Delta M = 3.29\), 95\% CI [2.99, 3.59], \(p < 0.001\)), HA vs. MA (\(\Delta M = 1.57\), 95\% CI [1.27, 1.87], \(p < 0.001\)), and MA vs. LA (\(\Delta M = 1.72\), 95\% CI [1.4183, 2.0185], \(p < 0.001\)). 

\subsection{Experiment 2: Haptic Experience}
\paragraph{Procedure:} After completing Experiment 1, participants freely explored the simulation for a duration of 15 minutes to test our final hypothesis \textbf{H4}. This allowed participants to navigate and interact with the environment at their discretion, including the option to surf, while being seated or kneeling on the physical board. Participants were asked to choose their preference of still water (no ocean ripples, i.e. no rocking board movement on water) or water with ocean ripples (board gently rocking), which remained consistent throughout the duration of their VR experience. The VR environment offered several interactive elements, such as a navigable half-submerged pirate ship,
fish, 
dolphins, and colorful poles serving as spatial references. These elements were collision-enabled, providing force feedback through the motion platform upon impact. The free exploration phase finished at a wave park that generated appropriately large waves for active surfing or a passive wave interaction, such as letting the wave rock the surfboard as it passes underneath. After the session, participants documented their haptic experience (HX) using the 5-point Likert scale questionnaire from~\cite{Anwar2023FactorsModalities}. They also completed a Simulator Sickness Questionnaire (SSQ)~\cite{kennedy1993simulator} twice -- once before Experiment 1 (pre) and once after completing Experiment 2 (post).

\paragraph{Results:} Six out of seventeen participants preferred navigating on still water, but all remarked that the ocean ripples increased realism and immersiveness.
The HX questionnaire can be grouped into four factors -- \textit{Realism} (convincing, believable haptic sensory alignment), \textit{Harmony} (seamless haptic-system-user integration), \textit{Involvement} (focused and meaningful user engagement) and \textit{Expressivity} (dynamic, responsive haptic feedback variation). The overall mean score across seventeen participants indicated our system to be fairly Realistic \(M=3.81 (SD = 0.87)\), achieved Harmony as \(M=3.67 (SD = 0.622)\), with high levels of Involvement \(M=4.41 (SD = 0.65)\), and Expressivity \(M=3.99 (SD = 0.72)\). Table \ref{Table1} lists the detailed HX questionnaire results.
The change in SSQ score (post$-$pre), across the full duration of the experiment, averaged across all 16 symptoms and 17 participants, was $M=0.39 (SD=0.42)$, with the highest recorded score post-VR exposure SSQ being 1.82, averaged across all symptoms.

\begin{table}[t!]
\caption{Results from Factors of HX Questionnaire}\label{Table1}
\begin{tabularx}{\textwidth}{|X|c|c|}
\hline
\textbf{Question: The haptic feedback …}             & \textbf{Mean (SD)} & \textbf{Factors}                                       \\ \hline
\rowcolor[HTML]{FFFFFF} 
Felt realistic (+)                                     & 3.65 (0.86)        & \cellcolor[HTML]{FFE5E3}                               \\ \cline{1-2}
\rowcolor[HTML]{FFFFFF} 
Was believable (+)                                    & 3.65 (0.99)        & \cellcolor[HTML]{FFE5E3}                               \\ \cline{1-2}
\rowcolor[HTML]{FFFFFF} 
Was convincing (+)                                       & 4.12 (0.70)        & \multirow{-3}{*}{\cellcolor[HTML]{FFE5E3}Realism}      \\ \hline
\rowcolor[HTML]{FFFFFF} 
Felt disconnected from the rest of the experience (-)    & 1.41 (0.71)        & \cellcolor[HTML]{DAFFDA}                               \\ \cline{1-2}
\rowcolor[HTML]{FFFFFF} 
Felt out of place (-)                                    & 1.41 (0.71)        & \cellcolor[HTML]{DAFFDA}                               \\ \cline{1-2}
\rowcolor[HTML]{FFFFFF} 
Distracted me from the task (-)                          & 1.18 (0.39)        & \multirow{-3}{*}{\cellcolor[HTML]{DAFFDA}Harmony}      \\ \hline
\rowcolor[HTML]{FFFFFF} 
Enjoyable as part of the experience (+)                  & 4.47 (0.72)        & \cellcolor[HTML]{DCDEFF}                               \\ \cline{1-2}
\rowcolor[HTML]{FFFFFF} 
Felt engaging with the system (+)                        & 4.35 (0.61)        & \multirow{-2}{*}{\cellcolor[HTML]{DCDEFF}Involvement}  \\ \hline
\rowcolor[HTML]{FFFFFF} 
All felt the same (-)                                   & 1.29 (0.59)        & \cellcolor[HTML]{D8FFFE}                               \\ \cline{1-2}
\rowcolor[HTML]{FFFFFF} 
Changes depending on how things change in the system (+) & 4.00 (0.79)        & \cellcolor[HTML]{D8FFFE}                               \\ \cline{1-2}
\rowcolor[HTML]{FFFFFF} 
Reflects varying inputs and events (+)                   & 4.18 (0.73)        & \multirow{-3}{*}{\cellcolor[HTML]{D8FFFE}Expressivity} \\ \hline
\end{tabularx}
\end{table}
\paragraph{\textbf{Discussion}:}The findings from Experiments 1, 2 and Fig. \ref{responseFig} validate our proposed mapping algorithm.
The results of the factorial ANOVA indicate that perception of acceleration within our surfing simulation is significantly impacted by the actual levels of acceleration provided and the presence of ocean ripples, hence proving \textbf{H1 and H3}. Notably, body posture did not have a significant effect on this perception, disproving \textbf{H2}. This is consistent with previous studies on the effect of body posture on vection \cite{guterman2012influence}, in which vection was found to be consistent across different upright body postures, such as sitting and standing. Vection is a visually induced illusion of self-motion that occurs when users perceive themselves as moving due to visual motion cues \textit{only}, despite being physically stationary. Although our system includes some visual cues that might induce feelings of self-motion, we notably also include kinesthetic cues of motion which have not previously been studied for effects of body posture. This novel finding that body posture does not affect acceleration perception even when kinesthetic cues of motion are provided in addition to visual cues increases our knowledge of human perception of self-motion.

Participants were able to effectively differentiate between the three acceleration levels in all body postures, with the low and high acceleration levels being most accurate. The perception of medium acceleration was a bit more varied, particularly in the kneeling position. Interestingly, acceleration was perceived to be slightly amplified in the presence of ocean ripples, as shown by the mean values in Fig.~\ref{responseFig}. This phenomenon could be attributed to the rougher motion experienced on wavy waters, e.g., driving a boat on a river versus the ocean.
The amplification of perceived acceleration may be due to the synergistic effect of kinesthetic and visually coherent oscillation cues across all 6 degrees of freedom (ocean ripples). This observation aligns with and expands upon prior findings, which indicate that visually coherent oscillations, orthogonal to the principal motion direction, significantly amplify vection~\cite{nakamura2010additional}. To delineate the relative contributions of visual oscillation cues alone versus the combined effect of visual and kinesthetic cues, further empirical investigations are warranted.
Overall, the data reveals a progressive increase in the mean values of perceived acceleration across the three acceleration levels, with a consistent trend of slightly higher values in ocean ripple conditions. 

Results from the HX questionnaire indicate that kinesthetic sensations experienced while navigating in VR correlated closely with the virtual surfboard movements. In addition to the surfboard movements being convincing, responsiveness to user inputs made the simulation more immersive and enjoyable, thus confirming \textbf{H4}. Furthermore, the validity of the simulation is supported by the time that participants spent in VR. They completed a total duration of 40 minutes in our simulation, with the exception of two individuals who discontinued their participation slightly earlier, at approximately 33 minutes, due to the onset of mild symptoms of simulator sickness. Overall, the SSQ scores show that the experience did not induce feelings of simulator sickness in most participants, which is especially promising for our system given the length of the time participants spent in VR. Some participants remarked it to be the best and most realistic VR experience they have had so far. A notable aspect of their experience was a marked unawareness of time while engaged in VR, leading to a pleasantly surprising realization upon learning the actual duration of their session.

\section{Conclusion and Future Work}
In this paper, we developed and evaluated a system for VR surfing using a 6-DOF motion platform, focusing on the simulation of aquatic motions driven by user-generated propulsive forces. Our algorithmic approach maps surfboard kinematics to motion outputs, validated by an experiment with 17 participants. Results confirm the system's ability to distinguishably reproduce various acceleration levels. The results also indicate that while body posture does not significantly impact acceleration perception, the introduction of ocean ripples perceptibly amplifies it. Additionally, our findings from the HX and SSQ questionnaires corroborate our approach of generating realistic motion cues in enhancing user engagement and perception. 
This study opens avenues for diverse applications, ranging from recreational and training purposes to therapeutic uses. It also presents a potential medium for surf therapy to enhance physical fitness, psychological well-being, and social interaction. This benefits groups such as at-risk youth and individuals with disabilities, including but not limited to those with chronic back pain and veterans, who lack access to actual surfing.

Future work will focus on enhancing realism and interactivity by incorporating pop-up maneuvers and force sensors to monitor pressure and balance, variable wind feedback, and considering further refinements in paddling techniques. 
The VR system will also be evaluated with surf therapy patients, comparing their virtual and real surfing experiences, thereby advancing the potential of this technology in therapeutic applications.

\printbibliography

@book{LeeVehicleEnvironments,
    title = {{Vehicle Simulation: Perceptual Fidelity in the Design of Virtual Environments}},
    author = {Lee, Alfred T},
    publisher = {CRC Press},
    year = {2017}
}

@article{SalterAutomatedVehicles,
    title = {{Automated Vehicles}},
    journal = {Int. J. of Mech. and Production Engineering},
    author = {Salter, S and Thake, D and Diels, S and Salter, Spencer and Mcsfs, Fimeche and Diels, Cyriel and Kanarachos, Stratis and Thake, Doug},
    isbn = {0711810117},
    issn = {2321-2071}
}

@inproceedings{Anwar2023FactorsModalities,
    title = {{Factors of Haptic Experience across Multiple Haptic Modalities}},
    year = {2023},
    author = {Anwar, Ahmed and Shi, Tianzheng and Schneider, Oliver},
    booktitle = {ACM CHI Conf. on Human Factors in Comp. Sys.},
    isbn = {9781450394215}
}

@article{good2005stopping,
  title={Stopping Distance and Acceleration Performance of Personal Watercraft},
  author={Good, Craig A and Paulo, Marshal H and Unger, Lonnie J and Varga, Janine L and Ellis, Mike C},
  journal={SAE transactions},
  year={2005},
  publisher={JSTOR}
}

@article{casas2017four,
  title={Four Different Multimodal Setups for Non-aerial Vehicle Simulations—a Case Study with a Speedboat Simulator},
  author={Casas, Sergio and Fern{\'a}ndez, Marcos and Riera, Jos{\'e} V},
  journal={Multimodal Technologies and Interaction},
  year={2017},
  publisher={MDPI}
}

@inproceedings{casas2013characterization,
  title={On the Characterization of a Speed-boat Motion for Real-time Motion Cueing},
  author={Casas, Sergio and Coma, Inmaculada and Riera, Jos{\'e} V and Fern{\'a}ndez, Marcos},
  booktitle={International Conference on Computer Graphics Theory and Applications},
  year={2013},
}

@article{nessler2019electromyographic,
  title={Electromyographic Analysis of the Surf Paddling Stroke across Multiple Intensities},
  author={Nessler, Jeff A and Ponce-Gonzalez, Jesus G and Robles-Rodriguez, Cristina and Furr, Heather and Warner, Mackenzie and Newcomer, Sean C},
  journal={The Journal of Strength \& Conditioning Research},
  year={2019},
  publisher={LWW}
}

@inproceedings{casas2012real,
  title={On the Real-time Physics Simulation of a Speed-boat Motion.},
  author={Casas, Sergio and Rueda, Silvia and Riera, Jos{\'e} V and Fern{\'a}ndez, Marcos},
  booktitle={GRAPP/IVAPP},
  year={2012}
}

@article{huang2023surfing,
  title={Surfing in Virtual Reality: An Application of Extended Technology Acceptance Model with Flow Theory},
  author={Huang, Yu-Chih and Li, Ling-Ni and Lee, Hsiao-Yun and Browning, Matthew HEM and Yu, Chia-Pin},
  journal={Computers in Human Behavior Reports},
  year={2023},
  publisher={Elsevier}
}

@article{ueng2008ship,
  title={A Ship Motion Simulation System},
  author={Ueng, Shyh-Kuang and Lin, David and Liu, Chieh-Hong},
  journal={VR},
  year={2008},
  publisher={Springer}
}

@inproceedings{mulder2012effect,
  title={The Effect of Motion on Presence during Virtual Sailing for Advanced Training},
  author={Mulder, Fabian A and Verlinden, Jouke C and Dukalski, Radoslaw R},
  booktitle={ISPR},
  year={2012}
}

@article{mulder2013development,
  title={Development of a Motion System for an Advanced Sailing Simulator},
  author={Mulder, Fabian A and Verlinden, Jouke C},
  journal={Procedia Engineering},
  year={2013},
  publisher={Elsevier}
}

@article{riera2022case,
  title={A Case Study on Vestibular Sensations in Driving Simulators},
  author={Riera, Jose V and Casas, Sergio and Alonso, Francisco and Fern{\'a}ndez, Marcos},
  journal={Sensors},
  year={2022},
  publisher={MDPI}
}

@article{von2008real,
  title={Real-time Rowing Simulator with Multimodal Feedback},
  author={von Zitzewitz, Joachim and Wolf, Peter and Novakovi{\'c}, Vladimir and Wellner, Mathias and Rauter, Georg and Brunschweiler, Andreas and Riener, Robert},
  journal={Sports Technology},
  year={2008},
  publisher={Wiley Online Library}
}

@article{liao2004novel,
  title={A novel washout filter design for a six degree-of-freedom motion simulator},
  author={Liao, Chung-Shu and Huang, Chih-Fang and Chieng, Wei-Hua},
  journal={JSME International Journal Series C Mechanical Systems, Machine Elements and Manufacturing},
  year={2004},
  publisher={The Japan Society of Mechanical Engineers}
}

@article{benninger2020surf,
  title={Surf therapy: A scoping review of the qualitative and quantitative research evidence},
  author={Benninger, Elizabeth and Curtis, Chloe and Sarkisian, Gregor V and Rogers, Carly M and Bender, Kailey and Comer, Megan},
  journal={Glob. J. Community Psychol. Pract},
  year={2020}
}

@inproceedings{labedan2021virtual,
  title={Virtual Reality for Pilot Training: Study of Cardiac Activity.},
  author={Labedan, Patrice and Darodes-de-Tailly, Nicolas and Dehais, Fr{\'e}d{\'e}ric and Peysakhovich, Vsevolod},
  booktitle={VISIGRAPP (2: HUCAPP)},
  year={2021}
}

@inproceedings{colombet2009motion,
  title={Motion cueing strategies for driving simulators},
  author={Colombet, Florent and Kemeny, Andras and M{\'e}rienne, Fr{\'e}d{\'e}ric and Pere, Christian},
  booktitle={ASME World Conf. on Innovative Virtual Reality},
  year={2009}
}

@article{guterman2012influence,
  title={Influence of head orientation and viewpoint oscillation on linear vection},
  author={Guterman, Pearl S and Allison, Robert S and Palmisano, Stephen and Zacher, James E},
  journal={Vestibular Research},
  year={2012},
  publisher={IOS Press}
}

@article{nakamura2010additional,
  title={Additional oscillation can facilitate visually induced self-motion perception: The effects of its coherence and amplitude gradient},
  author={Nakamura, Shinji},
  journal={Perception},
  year={2010},
  publisher={SAGE Publications Sage UK: London, England}
}

@article{kennedy1993simulator,
  title={Simulator sickness questionnaire: An enhanced method for quantifying simulator sickness},
  author={Kennedy, Robert S and Lane, Norman E and Berbaum, Kevin S and Lilienthal, Michael G},
  journal={The international journal of aviation psychology},
  year={1993},
  publisher={Taylor \& Francis}
}






\end{document}